\documentclass{article} 

\usepackage[utf8]{inputenc}
\usepackage{biblatex} %Imports biblatex package
\usepackage{graphicx}%
\usepackage{amsmath,amssymb,amsfonts}%
\usepackage{amsthm}%
\usepackage{mathrsfs}%
\usepackage[title]{appendix}%
\usepackage{xcolor}%
\usepackage{textcomp}%
\usepackage{manyfoot}%
\usepackage{booktabs}%
\usepackage{algorithm}%
\usepackage{algorithmicx}%
\usepackage{algpseudocode}%
\usepackage{listings}%
\usepackage{url}

\usepackage{hyperref}
\usepackage{graphbox}
\usepackage{subcaption}
\usepackage{tabularx}
\usepackage{multirow}
\usepackage{mathtools}
\usepackage{comment}
\usepackage{makecell}
\usepackage{authblk}

\newcommand{\mean}[1]{\langle#1\rangle}

\renewcommand{\S}[1][1]{\mathbb{S}^{#1}}
\renewcommand{\H}[1][2]{\mathbb{H}^{#1}}

\begin{document}

\title{Random Hyperbolic Graphs with Arbitrary Mesoscale Structures}
% \author{Stefano Guarino, Davide Torre, Enrico Mastrostefano}

\author[1]{{Stefano} {Guarino}\thanks{stefano.guarino@cnr.it}}

\author[1,2,3]{{Davide} {Torre}\thanks{dtorre@luiss.it}}

\author[1]{{Enrico} {Mastrostefano}\thanks{enrico.mastrostefano@cnr.it}}

\affil[1]{{Istituto per le Applicazioni del Calcolo ``Mauro Picone'' (CNR-IAC)}, {{Via dei Taurini 19}, {Rome}, {00185}, {Italy}}}

\affil[2]{{Libera Università Internazionale degli Studi Sociali ``Guido Carli'' (LUISS)}, {{Viale Romania, 32}, {Rome}, {00197}, {Italy}}}

\affil[3]{{ISI Foundation (ISI)}, {{Via Chisola, 5}, {Turin}, {10126}, {Italy}}}

\date{}

\maketitle

\begin{abstract}  
Real-world networks exhibit universal structural properties such as sparsity, small-worldness, heterogeneous degree distributions, high clustering, and community structures.
Geometric network models, particularly Random Hyperbolic Graphs (RHGs), effectively capture many of these features by embedding nodes in a latent similarity space.
However, networks are often characterized by specific connectivity patterns between groups of nodes --i.e. communities-- that are not geometric, in the sense that the dissimilarity between groups do not obey the triangle inequality.
Structuring connections only based on the interplay of similarity and popularity thus poses fundamental limitations on the mesoscale structure of the networks that RHGs can generate.
To address this limitation, we introduce the Random Hyperbolic Block Model (RHBM), which extends RHGs by incorporating block structures within a maximum-entropy framework.
We demonstrate the advantages of the RHBM through synthetic network analyses, highlighting its ability to preserve community structures where purely geometric models fail.
Our findings emphasize the importance of latent geometry in network modeling while addressing its limitations in controlling mesoscale mixing patterns.
\end{abstract}

% \section{Introduction and Background}

\vspace{1cm}

\section{Introduction}

% As researchers seek to gain deeper insights into the basic mechanisms governing real-world network formation, 
% have emerged that 
Models of complex networks aim to replicate structural patterns observed in real-world systems while enabling effective network randomization for benchmarking and hypothesis testing~\cite{archana2014community, salathe2010dynamics, celestini2021epidemics}.
Over the past few decades, numerous studies have identified a set of universal characteristics that most real-world networks share.
These include sparsity~\cite{del2011all}, the small-world property~\cite{watts1998collective}, an inhomogeneous degree distribution~\cite{barabasi1999emergence} and a high clustering coefficient~\cite{krioukov2016clustering}.
Most real networks, however, also exhibit non-trivial mesoscale structures, such as communities with specific mixing patterns~\cite{toivonen2006model}.
Defining network models that provide explicit control of statistics at all scales is crucial for realistic network representations~\cite{girvan2002community, cherifi2019community}.

%various network models have emerged, each aiming to reflect and potentially explain specific observed features of complex networks~\cite{newman2018networks}. However, formulating sound and general models that encompass all of these features remains a significant theoretical challenge.

Random Hyperbolic Graphs (RHG) provide a compelling framework for explaining various observed network characteristics by embedding nodes in a hidden metric space with negative curvature~\cite{krioukov2010hyperbolic, zuev2015emergence, krioukov2016clustering, Kovacs2021inherent, serrano2022shortest}.  
Such a latent geometry provides a powerful approach to modeling networks, structuring connections based on both similarity and popularity~\cite{boguna2010sustaining, muscoloni2019angular, yang2020high}.  
In the \(\mathbb{S}^D\) model, node popularity is controlled by latent degrees \(\kappa_i\) and similarity by node-to-node distances \(x_{ij} = R\Delta\theta_{ij}\) within a \(D\)-dimensional sphere.
The edge probability reads  
\begin{equation}\label{eq:p_S1}
    p_{ij}= \frac{1}{1+\left(\frac{x_{ij}}{\left(\mu\kappa_i\kappa_j \right)^{1/D}}\right)^\beta}.
\end{equation}  
where the inverse temperature \(\beta\) controls the clustering coefficient, whereas the density parameter \(\mu\)
%= \frac{R \beta \sin(\pi/\beta)}{N\mean{\kappa}}\) 
ensures the correct average degree.
% This functional form arises from a maximum entropy principle where constraining both the degree sequence and total system energy, by setting \(\varepsilon(x_{ij})=\ln(x_{ij})\) as the link energy—minimizes non-structural degree correlations~\cite{boguna2020small}.  
The hidden degrees in \(\mathbb{S}^D\) can be mapped onto radial coordinates in a hyperbolic plane \(\mathbb{H}^{D+1}\), transforming the spherical representation with hidden degrees into a purely geometric hyperbolic one.
When hidden degrees follow a power-law distribution with exponent \(\gamma \geq 2\), the \(\mathbb{S}^D\) and \(\mathbb{H}^{D+1}\) models become isomorphic in the thermodynamic limit~\cite{krioukov2010hyperbolic}.  
The popularity-similarity optimization (PSO) model~\cite{papadopoulos2012popularity} and its nonuniform generalization (nPSO)~\cite{muscoloni2018nonuniform} extend the ideas of RHG by incorporating dynamic elements that reflect the balance between popularity and similarity in the emergence of real networks.
Despite the conceptual elegance of hyperbolic spaces and the potential benefits of higher dimensions (\(D>1\)), the \(\mathbb{S}^1\) model remains the most widely used due to its simplicity. It is characterized by two fundamental properties: (i) when angular coordinates are uniformly distributed, the expected degree of a vertex converges to its hidden degree in the limit of large networks, \(\mean{\deg_i}_{\S} \to \kappa_i\); and (ii) clustering decreases as \(\beta \to 1\) and stabilizes at a constant for \(\beta \to +\infty\).

Hyperbolic latent spaces naturally capture hierarchical and transitive structures, mirroring empirical properties such as degree heterogeneity and clustering~\cite{garcia2018soft, zuev2015emergence, muscoloni2018nonuniform}.
A fundamental limitation of these models, however, is their lack of explicit control over mixing patterns at intermediate scales.
Both RHG and PSO models can produce networks with strong communities~\cite{Kovacs2021inherent, wang2016hyperbolic, faqeeh2018characterizing}, but these arise implicitly rather than as a controlled feature. 
Modularity can be imposed via angular node aggregation~\cite{boguna2010sustaining, garcia2018soft, muscoloni2018nonuniform, jankowski2023dmercator}, but inter-community connectivity is entirely driven by similarity in the latent space.
This limitation hinders the application of these models in all cases where mesoscale structures arise from shared categorical attributes within groups, such as interests in social networks or topics in information networks~\cite{faqeeh2018characterizing, bruno2019community, wang2016hyperbolic, ye2022hyperbolic}.
Even scalar attributes often act like categorical ones, with connectivity in the network not reflecting proximity in the underlying metric space --think, for instance, to strong sub-diagonals in age-based mixing matrices, or to co-expression networks where modular patterns are mostly driven by pathway membership rather than proximity in expression space.
Increasing the dimensionality of latent hyperbolic spaces has been suggested as a way to enable finer control over group interactions and overcome the intrinsic correlations between node distances found in low dimensions~\cite{Almagro2022detecting, kovacs2022generalised, budel2024random, desy2023dimension}.
However, existing models lack mechanisms for leveraging dimensionality to enforce specific community mixing patterns while preserving essential network properties.
More generally, whether arbitrary group-mixing patterns are compatible with a popularity-similarity framework is yet to be clarified in the literature.
% More generally, whether arbitrary mesoscale structures are compatible with a popularity-similarity framework is yet to be clarified in the literature.
% Although geometric models provide direct control over degree distributions and transitivity, they lack explicit mechanisms to regulate community structures and mixing patterns. 

In this paper, we propose the \textit{Random Hyperbolic Block Model} (RHBM),
%\footnote{\color{blue}[CAMBIARE NOME? NON DEFINIAMO IL MODELLO IPERBOLICO NÉ MOSTRIAMO L'EQUIVALENZA]},
a generalization of RHGs that incorporates explicit control over the network's mixing matrix by means of a set of hidden forces binding node groups.
Instead of hidden degrees, we introduce a hidden fitness parameter, normalized according to block connectivity. 
We show that the RHBM, as the $\S[1]$ model~\cite{boguna2020small}, can be interpreted as an entropy-maximizing probabilistic mixture of grand canonical network ensembles.
% , ,  constrained by expected total energy, degree sequence and number of edges between each pair of groups.
% --as in the $\S[1]$ model-- 
% This connection to entropy maximization suggests a pathway to extending RHGs with additional constraints, leading to exponential random graphs embedded in a latent similarity space. 
This means that the proposed formulation is optimal for geometric networks with constrained expected degree sequence and group-mixing matrix.
% More generally, the max-entropy approach suggests a pathway to extending RHGs with additional constraints,  exponential random graphs embedded in a latent similarity space.
% when the edge probability follows a Fermi–Dirac distribution, as in (\ref{eq:p_S1}),
We demonstrate that the RHBM can be equivalently represented as the union of \(\binom{n}{2}\) \(\S\) graphs, corresponding to intra- and inter-block connections among \(n\) communities, provided that node latent features are drawn once and applied consistently across subgraphs, ensuring stable centrality and homophily patterns.
Due to this equivalence, we introduce the alternative notation \((\S)^{\binom{n}{2}}\) for this graph model.
% The RHBM builds on that idea to extend both the degree-corrected stochastic block model and the \(\S\) model, preserving a combination of their properties.

To highlight the relevance of the presented model, we present experimental results in support of the intuition that attribute-based group-mixing is, in general, non-geometric.
We generated synthetic networks using the RHBM and embedded them in $\S[D]$, for $D=1,\ldots,5$, using the state-of-the-art tool D-Mercator~\cite{jankowski2023dmercator}.
In D-Mercator, the algorithm infers the inverse temperature $\beta$, the parameter $\mu$ controlling the network's average degree, the radius of the $\S[D]$ disk, the hidden degrees and the latent angular coordinates, maximizing the likelihood of the inferred latent coordinates given the observed graph. 
By analyzing the graphs obtained from the embedding, we show that models based solely on the popularity-similarity paradigm can capture local properties such as degree distributions and clustering but fail to enforce specific mesoscale structures like community mixing.
Even the choice of model parameters and latent coordinates that best explains the input network, in fact, fails to explain the input mixing patterns, regardless of \(D\).  
% We also compare real-world networks against randomized versions generated by RHBM and by sampling from D-Mercator embeddings, assessing which structural properties are better preserved by each approach.
In summary, we show that our proposed RHBM, by explicitly incorporating block-level constraints, permits to regulate community structures and mixing patterns while maintaining the fundamental strengths of hyperbolic network models.
Our findings challenge the notion that purely geometric models suffice to explain real-world network structures. 
% Our approach introduces an additional term in the edge probability \( p_{ij} \) to overcome this limitation, and we discuss how it generalizes to higher-dimensional latent spaces \( D > 1 \).

% analytical arguments in support of the distinction between attribute-based group-mixing and individual popularity/similarity-based mixing, following the rationale that the triangle inequality in a suitable similarity metric space explains high clustering, but poses fundamental limitations to block-wise connectivity patterns.

\section{The model}

% Geometric models rely on the intuition that the topological structure of real complex networks is compatible with an underlying latent metric space, that describes the combined effect of two main drivers of edge creation: popularity and similarity. 

% \subsection*{The Random Hyperbolic Block Model}
The Random Hyperbolic Block Model (RHBM) extends the $\S$/$\H$ models by addressing the case in which the $N$ nodes are organized into $n$ disjoint \textit{blocks} (or \textit{communities}), whose connectivity patterns are controlled by a symmetric matrix $\Phi=\{\Phi_{IJ}\}$ of latent forces\footnote{In the following, we use lower-case indices for nodes and upper-case indices for blocks, with $i\in I$ indicating that node $i$ belongs to block $I$, and $I_i$ denoting $i$'s block.}.
As in the $\S$ model, the nodes are placed uniformly at random on a disk of radius $R$ by assigning angular coordinate $\theta_i\in[0,2\pi)$ to node $i$.
Instead of a hidden degree, each node has a (hidden) fitness $\phi_i$ drawn from a suitable distribution --in the following, we will assume power-law distributed fitness with exponent $\gamma=2.5$, but any distribution is potentially usable. 
The probability per link reads
\begin{equation}\label{eq:p_ij}
p_{ij}=\frac{1}{1+\left(\frac{x_{ij}}{\tilde{\mu}_{IJ} \phi_i \phi_j \Phi_{IJ}} \right)^{\beta}} 
\end{equation}
where $i\in I$, $j\in J$, $x_{ij}=R\Delta\theta_{ij}$ and $\tilde{\mu}_{IJ}=R \beta \sin(\pi/\beta)$.
If we denote $\kappa_{iJ}=\phi_i \Phi_{IJ}$ the expected degree of $i$ towards block $J$ --i.e. the number of nodes in block $J$ to which $i$ is connected-- and $\mu_{IJ}=\frac{\tilde{\mu}_{IJ}}{\Phi_{IJ}}$, \eqref{eq:p_ij} can be rewritten as
\begin{equation}\label{eq:p_ij_other}
p_{ij}=\frac{1}{1+\left(\frac{x_{ij}}{\kappa_{iJ} \kappa_{jI} \mu_{IJ}} \right)^{\beta}} 
\end{equation}
This shows that in the RHBM $p_{ij}$ takes the same functional form of (\ref{eq:p_S1}), with the only difference that the total hidden degrees $\kappa_i,\kappa_j$ and the normalization constant $\mu$ are replaced by their block-wise counterparts $\kappa_{iJ},\kappa_{jI}$ and $\mu_{IJ}$.

\subsection*{Maximum entropy formulation}
In the thermodynamic limit $N\to \infty$, the RHBM can be equivalently defined as the maximum-entropy ensemble $P$ satisfying 
\begin{align}
    \label{eq:HGBM_energy}
    \mean{E}_P &= W \\
    \label{eq:HGBM_blocks}
    \mean{L_{IJ}}_P &= F_{IJ} \quad \text{for all } I, J\\  %p\binom{N}{2}\Delta_{IJ}
    \label{eq:HGBM_degree}
    \mean{\deg_i}_P &= f_i \sum_J F_{I_iJ} \quad \text{for all } i   %p\binom{N}{2} \Delta_{I_iJ} 
\end{align}
% where $\mean{\cdot}_P$ denotes the expectation with respect to $P$.
Equation (\ref{eq:HGBM_energy}) binds the expected total energy $E=\sum_{ij} \varepsilon(x_{ij})$ to $W$, (\ref{eq:HGBM_blocks}) sets the expected number of links between blocks $I$ and $J$ (or twice that number if $I=J$) to $F_{IJ}$, and (\ref{eq:HGBM_degree}) controls the expected degree sequence of the network.
The consistency of (\ref{eq:HGBM_blocks}) and (\ref{eq:HGBM_degree}) is guaranteed by taking $f_i$ so that $\sum_{i\in I} f_i = 1$ for all $I$.
$f_i$ can thus be interpreted as the fraction of all edges incident on block $I$ that are incident on node $i$. %, with $\sum_{i\in I} f_i = 1$ for all $I$.

As for the standard $\S$ model, the maximum entropy probability per graph factorises in terms of independent probabilities per link as
\[  
p_{ij}=\frac{1}{1+e^{\lambda_i+\lambda_j+\eta_{IJ}+\beta\varepsilon(x_{ij})}} % \quad \text{if } i\neq j \text{, 0 otherwise}
%\dfrac{x_ix_jy_{I_iJ_j}}{1+x_ix_jy_{I_iJ_j}} \quad \text{if } i\neq j \text{, 0 otherwise}
\]
where $i\in I$, $j\in J$, and $\beta$, $\eta_{IJ}$ and $\lambda_i$ are the Lagrange multipliers associated, respectively, to conditions~\ref{eq:HGBM_energy},~\ref{eq:HGBM_blocks} and~\ref{eq:HGBM_degree}. 
Also, with the same line of reasoning used in~\cite{boguna2020small}, it can be proved that taking $\varepsilon(x_{ij})=\ln(x_{ij})$ is the only way to suppress nonstructural degree correlations.
This leads to
\begin{equation}\label{eq:p_ent}
p_{ij}=\frac{1}{1+\left(x_{ij} e^{\frac{\lambda_i+\lambda_j+\eta_{IJ}}{\beta}}\right)^{\beta}} 
%\dfrac{x_ix_jy_{I_iJ_j}}{1+x_ix_jy_{I_iJ_j}} \quad \text{if } i\neq j \text{, 0 otherwise}
\end{equation}
% for all $i\neq j$, $p_{ii}=0$ for all $i$.

For fixed $\lambda_i$ and $\eta_{IJ}$ (i.e., fixed $f_i$ and $F_{IJ}$), the expected probability that nodes $i$ and $j$ are connected can be found integrating $p_{ij}$ over $\theta_j$ and $\lambda_j$.
In the limit $N\to \infty$,\footnote{Eq. (\ref{eq:mean_p_ij}) follows from $\frac{1}{X} \int_0^{X} \frac{dt}{1+t^{\beta}} \stackrel{X\to\infty}{\approx} \frac{1}{X} \frac{\pi}{\beta \sin(\pi/\beta)}$.} one obtains
\begin{equation}\label{eq:mean_p_ij}
    \mean{p_{ij}\mid \lambda_i, \eta_{IJ}} = \frac{1}{\tilde{\mu}_{IJ}} e^{-\frac{\eta_{IJ}}{\beta}} e^{-\frac{\lambda_i}{\beta}} \mean{e^{-\frac{\lambda}{\beta}}}  
\end{equation}
Further integrating (\ref{eq:mean_p_ij}) over $\lambda_i$ gives
\begin{align}
\label{eq:eta}
e^{-\frac{\eta_{IJ}}{\beta}} &= \frac{\tilde{\mu}_{IJ} F_{IJ}}{N_I N_J} \frac{1}{\mean{e^{-\frac{\lambda}{\beta}}}^2}\\ 
\label{eq:mean_deg_i}
e^{-\frac{\lambda_i}{\beta}} &= f_i N_I \mean{e^{-\frac{\lambda}{\beta}}}
\end{align}
which, plugged into (\ref{eq:p_ent}), give
\begin{equation}\label{eq:p_final}
p_{ij}=\frac{1}{1+\left(\frac{x_{ij}}{\tilde{\mu}_{IJ} f_i f_j F_{IJ}}\right)^{\beta}} 
\end{equation}
Equation (\ref{eq:p_final}) shows that this maximum-entropy approach yields exactly the model defined by (\ref{eq:p_ij}), provided that $\phi_i=f_i$ and $\Phi_{IJ}=F_{IJ}$ --i.e., that the latent fitness equals the block-normalized degree, and that the hidden force between two blocks equals the expected number of edges between them.
Let us underline that the equivalence works with exactly $\phi_i=f_i$ and $\Phi_{IJ}=F_{IJ}$ only in the thermodynamic limit $N\to \infty$.
In a finite system, the values for $\phi_i$ and $\Phi_{IJ}$ that satisfy \eqref{eq:HGBM_blocks} and \eqref{eq:HGBM_degree} for given $f_i$ and $F_{IJ}$ must be found numerically.

\subsection*{Alternative formulation: $(\S)^{\binom{n}{2}}$}
The RHBM can be equivalently obtained under the following system of \emph{stronger} constraints~\cite{guarino2023hidden}
\begin{align}
    \label{eq:HGBM_star_energy2}
    \mean{E}_P &= W \\
    % \label{eq:HGBM_blocks2}
    % \mean{L_{IJ}}_P &= K_{IJ} \quad \text{for all } I, J\\
    \label{eq:HGBM_star_degree2}
    \mean{\deg_{iJ}}_P &= f_i F_{IJ} \quad \text{for all } i,J
\end{align}
where $\deg_{iJ}$ is the degree of $i$ towards $J$.
Equation (\ref{eq:HGBM_star_degree2}), which implies both (\ref{eq:HGBM_blocks}) and (\ref{eq:HGBM_degree}), states that $f_i$, instead of just controlling $i$'s expected total degree, controls $i$'s expected degree towards each and every blocks in the graph.
The inverse --(\ref{eq:HGBM_blocks}) and (\ref{eq:HGBM_degree}) implying (\ref{eq:HGBM_star_degree2})-- is not true in general, but it is true under the maximum entropy principle, because the least restrictive assumption one can make is that the popularity of a node does not vary from one block to another.
It is easy to verify that a model defined in terms of (\ref{eq:HGBM_star_energy2}) and (\ref{eq:HGBM_star_degree2}) is equivalent to the union of the $\binom{n}{2}$ graphs obtained imposing 
\begin{align}
    \label{eq:HGBM_energy3}
    \mean{E_{IJ}}_P &= W_{IJ}\\
    % \label{eq:HGBM_blocks3}
    % \mean{L_{IJ}}_P &= K_{IJ} \quad \text{for all } I, J\\
    \label{eq:HGBM_degree3.1}
    \mean{\deg_{iJ}}_P &= f_i F_{IJ} \quad \text{for all } i\in I\\
    \label{eq:HGBM_degree3.2}
    \mean{\deg_{jI}}_P &= f_j F_{IJ} \quad \text{for all } j\in J
\end{align}
for all possible choices of $I,J$ and suitable $W_{IJ}$, where conditions (\ref{eq:HGBM_energy3}), (\ref{eq:HGBM_degree3.1}) and (\ref{eq:HGBM_degree3.2}) define a bipartite graph if $I\neq J$.
In particular, this implies that the $\S$ model can be generalized to accommodate a community structure by simply considering each subgraph defined by a pair $(I,J)$ as an independent mono/bipartite $\S$ graph, provided that: (i) the hidden degrees are replaced with block-normalized hidden fitnesses; (ii) both the hidden fitness $\phi_i$ and the latent angular coordinate $\theta_i$ associated to vertex $i$ are drawn just once and for all --i.e., not re-drawn for each subgraph.
The fact that node $i$ has the same $\phi_i$ in all subgraphs guarantees that (\ref{eq:HGBM_degree}) holds and that the graph has, in good approximation, the desired degree distribution.
On the other hand, the fact that $i$ has the same $\theta_i$ in all subgraphs preserves the transitivity of the vertex similarity across different subgraphs, so as to guarantees the desired high clustering for the entire network.

\section{Experiments}

To assess the expressive power and limitations of the RHBM, we conducted a series of controlled experiments, aimed at testing how well a purely geometric embedding—specifically, the one produced by D-Mercator—can recover key structural properties of synthetic networks generated with the model.

\subsection{Experimental Setup}

The base parameters of the RHBM are:
\begin{itemize}
    \item $N$: number of nodes
    \item $k$: average degree
    \item $\gamma$: exponent of the power-law degree/fitness distribution
    \item $\beta$: inverse temperature
    \item $n$: number of communities
\end{itemize}

We have set the mixing matrix $F$ using a parametric model that allows us to consider a variable range of mixing patterns:
\begin{equation*}
    F = \frac{\rho+1}{n} \begin{bmatrix}
1 & 0 & 0 & \cdots & 0 \\
0 & 1 & 0 & \cdots & 0 \\
0 & 0 & 1 & \cdots & 0 \\
\vdots& & & \ddots &   \\
0 & 0 & 0 & \cdots & 1
\end{bmatrix} + \frac{1-\rho}{2\sum_{i=1}^n (n-i)q^i} \begin{bmatrix}
0 & q & q^2 & \cdots & q^{n-1} \\
q & 0 & q & \cdots & q^{n-2} \\
q^2 & q & 0 & \cdots & q^{n-3} \\
\vdots& & & \ddots &   \\
q^{n-1} & q^{n-2} & q^{n-3} & \cdots & 0
\end{bmatrix}
\end{equation*}
The parameter $\rho \in [-1, 1]$ interpolates between disassortative and assortative mixing, while $q \in (0,1]$ controls the decay of connectivity away from the diagonal, making it possible to obtain ordered community structures.
% When $\rho = 0$, off-diagonal terms weighted equally; when $q = 1$, off-diagonal weights are flat.

We generated networks using the RHBM, varying one parameter at a time from the sets
    $N\in \{1000, 3000, 5000\}$, $k\in\{5, 10, 20\}$, $n\in\{2, 10, 100\}$, $\rho\in\{-0.5, 0, 0.5\}$, $q\in\{0.5, 0.75, 1\}$, $\beta\in\{2, 5, 10\}$, while keeping all others fixed at $N=3000$, $k=10$, $\gamma=2.5$, $n=10$, $\rho=0.5$, $q=1$ $\beta=2$.
Each generated network was embedded using D-Mercator, varying the latent space dimension \( D \in \{1, 2, 3, 4, 5\} \). From each embedding, we computed:
\begin{enumerate}
    \item The expected mixing matrix (based on the edge probability formula in the $\S[D]$ model)
    \item The expected degree sequence
    \item Ten synthetic networks sampled from the inferred edge probabilities, used to compute global and average local clustering coefficients
\end{enumerate}

\subsection{Results and Evaluation}

\begin{figure}[htbp]
    \centering
    \includegraphics[width=0.98\textwidth]{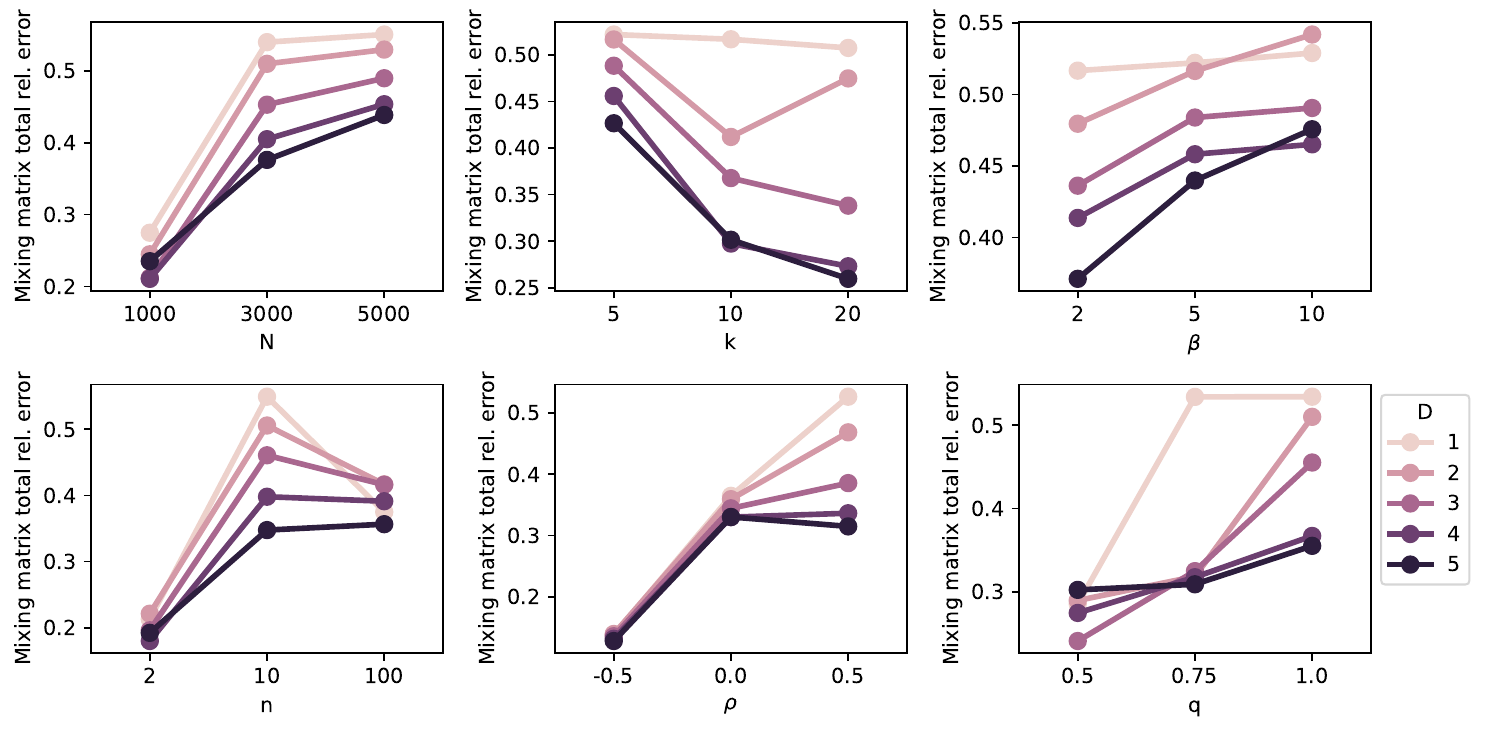}
    \caption{Relative error in recovering the mixing matrix for various model configurations and embedding dimensions.}
    \label{fig:mixing_error}
\end{figure}

\paragraph{Block Mixing Matrix Reconstruction.}  
To evaluate the accuracy of mixing matrix reconstruction, we computed the relative absolute error:
\[
\frac{|\!|F_{\text{out}} - F_{\text{in}}|\!|_1}{|\!|F_{\text{in}}|\!|_1}
\]
where \(F_{\text{in}}\) is the ground-truth input matrix and \(F_{\text{out}}\) the one expected from the embedding. The results (Figure~\ref{fig:mixing_error}) show that the error is generally significant, ranging from \(10\%\)–\(15\%\) up to \(50\%\)–\(55\%\). Errors tend to be lower for smaller networks, fewer communities, more disassortative or orderly mixing patterns, and higher embedding dimensions.

\begin{figure}[htbp]
    \centering
    \includegraphics[width=0.98\textwidth]{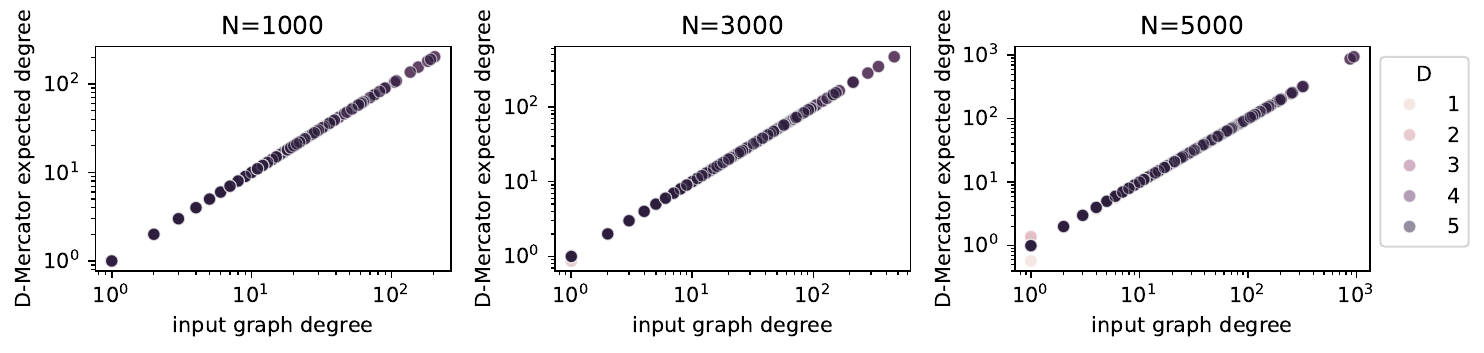}

    \includegraphics[width=0.98\textwidth]{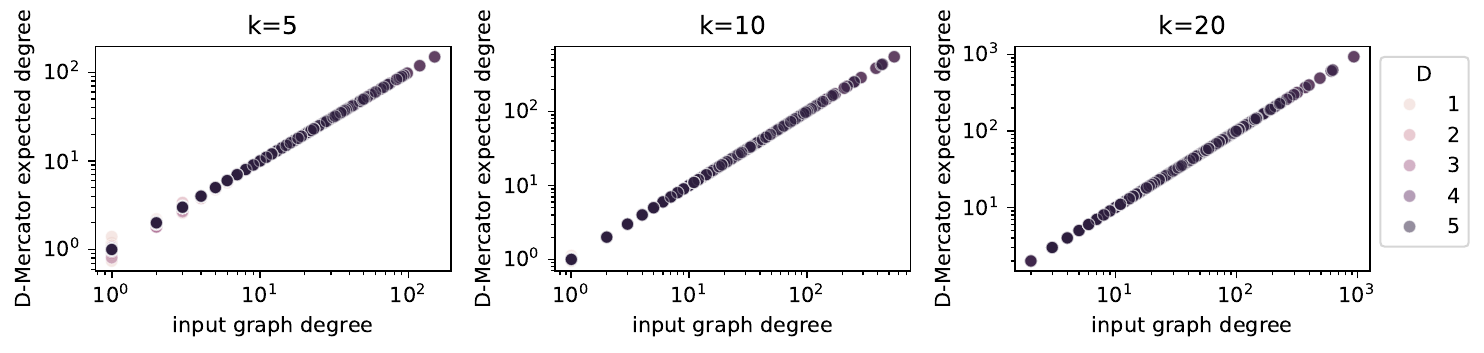}

    \includegraphics[width=0.98\textwidth]{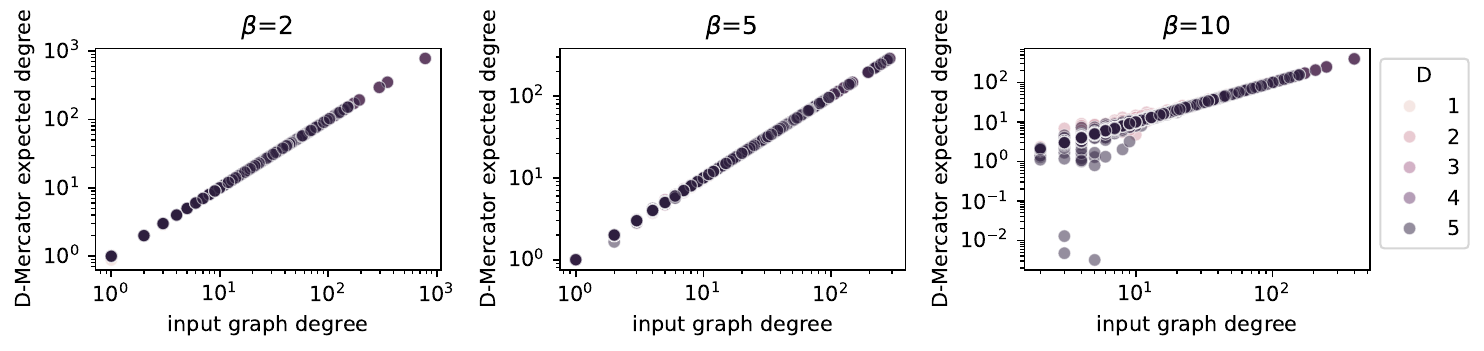}

    \includegraphics[width=0.98\textwidth]{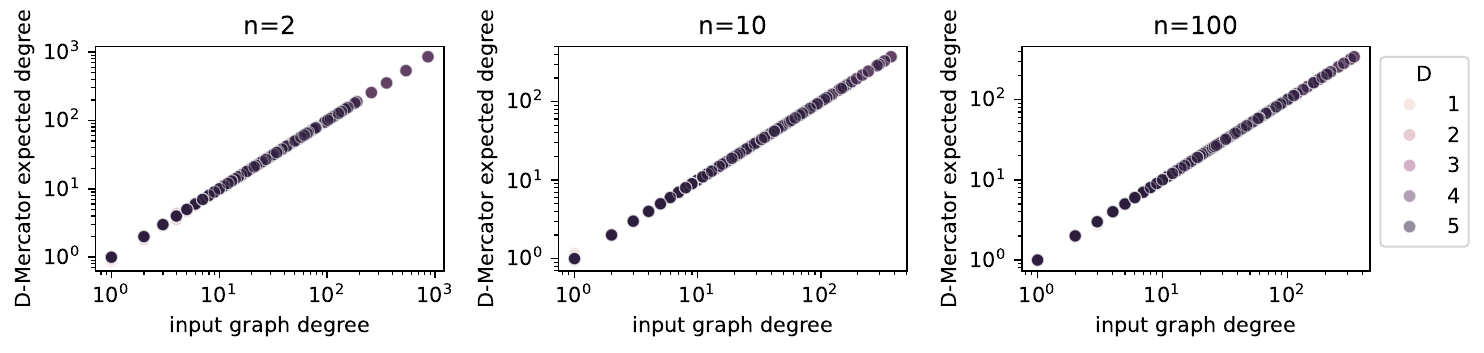}

    \includegraphics[width=0.98\textwidth]{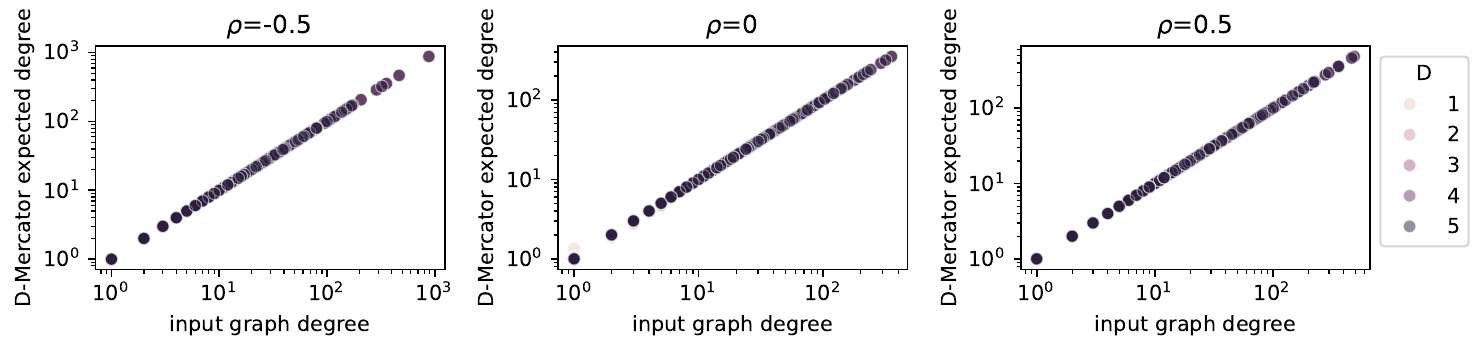}

    \includegraphics[width=0.98\textwidth]{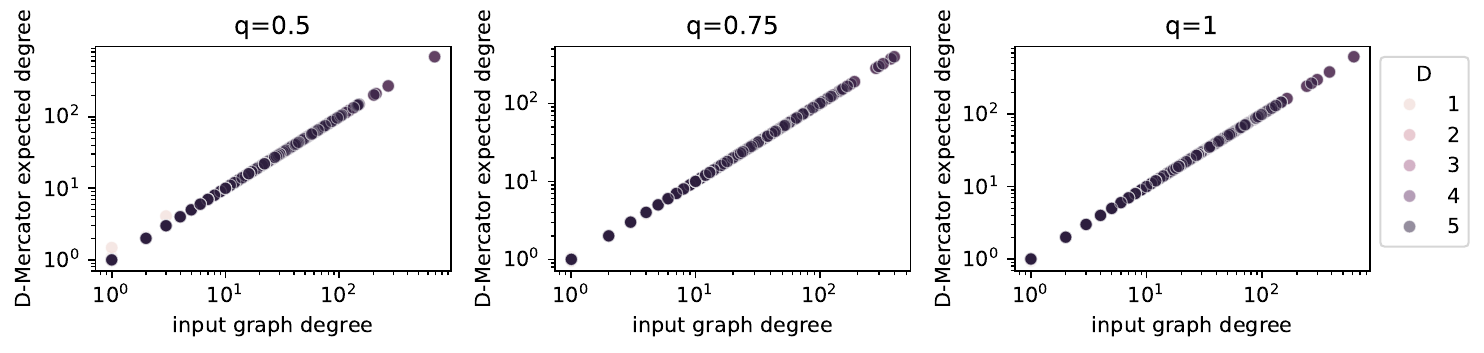}
    \caption{Scatter plot of input vs. reconstructed degree sequences. All points lie close to the identity line.}
    \label{fig:degree_scatter}
\end{figure}

\paragraph{Degree Sequence Reconstruction.}  
We compared the expected degree sequence derived from the embedding with the one of the input RHBM graph. As shown in Figure~\ref{fig:degree_scatter}, the match is nearly perfect across all tested configurations, confirming that D-Mercator captures node popularity extremely well.

\begin{figure}[htbp]
    \centering
    \includegraphics[width=0.98\textwidth]{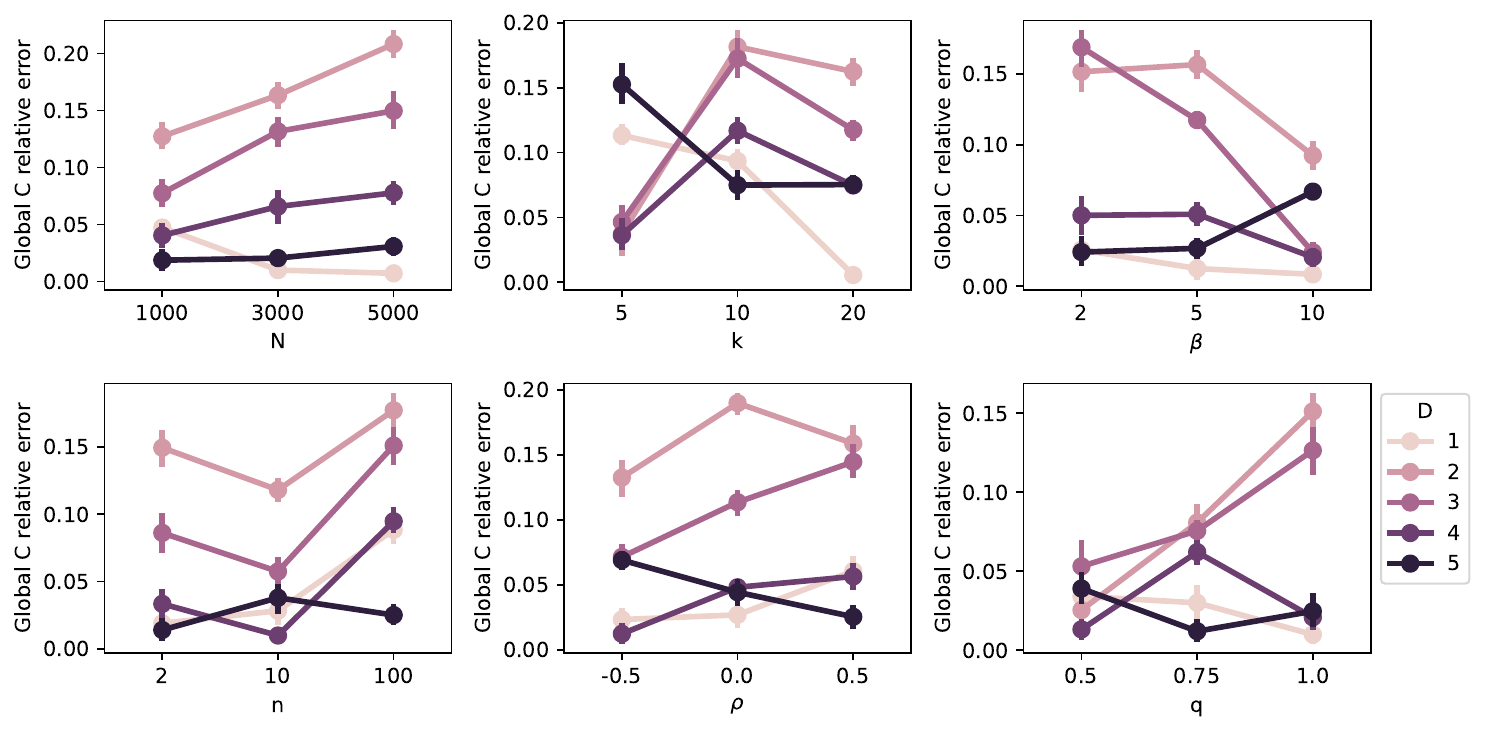}
    \caption{Relative error in global clustering coefficients. D-Mercator preserves clustering well across all configurations.}
    \label{fig:clustering_error_global}
\end{figure}

\begin{figure}[ht]
    \centering
    \includegraphics[width=0.98\textwidth]{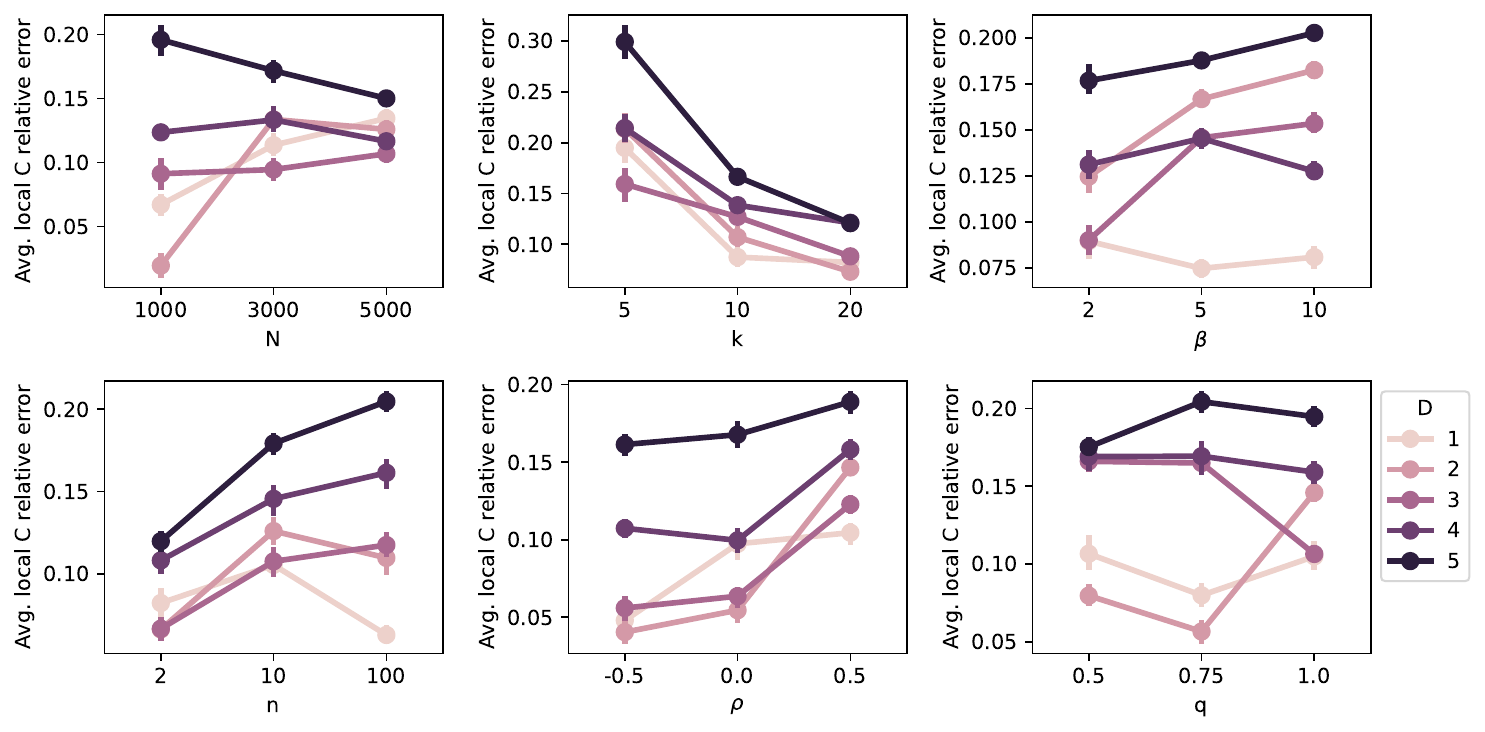}
    \caption{Relative error in average local clustering coefficients. Again, D-Mercator preserves clustering well across all configurations.}
    \label{fig:clustering_error_local}
\end{figure}

\paragraph{Clustering Coefficient.}  
We computed the relative error in both the global and average local clustering coefficients:
\[
\frac{|C - C_{\text{in}}|}{C_{\text{in}}}
\]
Figures~\ref{fig:clustering_error_global} and~\ref{fig:clustering_error_local} show that the error is consistently low—below \(20\%\)–\(30\%\) in all cases. Local clustering is best preserved for \(D=1\), while global clustering shows slightly better fidelity for \(D=1\) and \(D=5\), without major differences across parameter settings.

\subsection{Discussion}

Overall, the results suggest that D-Mercator embeddings of networks generated with the RHBM are highly effective at preserving local structural properties such as node degrees and clustering, even under diverse parameter regimes. However, they fall short of recovering mesoscopic structures such as group-to-group connectivity patterns, indicating that purely geometric representations may not suffice to explain intermediate-scale structure when group mixing is imposed independently of geometry.

\section{Conclusion}

Intermediate-scale structures are typical of real-world networks and significantly influence their function and dynamics.
While defining what constitutes a ``community'' in a network is often challenging, synthetic networks frequently need to replicate specific connection patterns between groups of nodes identified by attributes.
Similarly, when abstracting an observed graph by defining an appropriate null model that preserves its key properties, it is impossible to overlook the presence of groups of nodes that are particularly well-connected internally or to each other.  

In this article, we introduced the Random Hyperbolic Block Model, an extension of the $\S$ random geometric graph that allows for explicit control over the frequency of connections between different groups of nodes in the network.  
Like the $\S$ model, the RHBM can be derived through an entropy maximization principle, with the key difference that it includes an additional constraint on the expected mixing matrix between groups. This leads to an expression for edge probabilities that retains the same functional form as in the $\S$ model, but with hidden degrees and normalization constants replaced by block-specific counterparts that depend on the desired connection strength between communities. In doing so, the RHBM extends latent geometry models—which capture many real-world network features through a balance of popularity and similarity—by incorporating an explicit and tunable mechanism for controlling inter-group connectivity.

The common approach used in the literature to enforce a community structure in $\S[D]$ was using a different spatial distribution for each block of nodes~\cite{boguna2010sustaining, garcia2018soft, muscoloni2018nonuniform, jankowski2023dmercator}.
% While having block-dependent spatial densities seems like a necessary condition to obtain a specific partition of nodes into communities, it is not sufficient, in general, to gain complete control over inter-community connectivity.
% The latent geometry of the network guarantees strong clustering but at the expenses of freedom in the meso-scale structure of the network.
In particular, spatial aggregation of nodes in the same community leads to assortative mixing, where intra-block edges are more likely than inter-block ones.
% , and the variance of each block's distribution and the distances between their centroids provide some level of control on the density of links within communities with respect to those occurring between them.
Due to the triangle inequality, however, one cannot have strong connectivity between both pairs of blocks \((I,M)\) and \((M,J)\) without also having non-negligible connectivity between \((I,J)\), unless one allows very large variances that reduce the meaning of group structure.
In addition, the nodes being uniformly distributed in the latent space is one key assumption of the $\S[D]$ model, that guarantees node-homogeneity and makes the model analytically treatable.
% Variants of the model where this assumption is dropped have offered relevant insights, but did not fully unfold the impact of allowing variable density patterns on both community mixing and other structural properties of the network~\cite{garcia2018soft, muscoloni2018nonuniform}.
The RHBM addresses this issue decoupling attribute-based group-mixing and individual popularity/similarity-based mixing.

To highlight the importance of the proposed model, we conducted experiments to assess whether networks created with the RHBM have a more natural, fully geometric interpretation.
To this end, we fed graphs obtained with the RHBM to the D-Mercator algorithm --the state-of-the-art for multidimensional embeddings in \(\mathbb{S}^D\)-- varying the latent space dimension \(D\).
D-Mercator does not explicitly attempt at reconstructing the input attribute-based mixing, but it tries to identify the most suitable geometric representation for that network, given the observed connectivity patterns.
By letting the input network change in aspects like the number of communities and the type of mixing, we verified that even the most plausible geometric configuration is generally insufficient to fully explain or reproduce the input intermediate-scale patterns.
% Even if an alternative and more accurate geometric interpretation of the network existed, it would be suboptimal when compared with the RHBM guarantees that the network structure is maximally random except for the imposed constraints.
These results suggest that the RHBM offers a principled and flexible framework to model structured networks where community structure and geometric organization coexist, without artificially coupling the two.

Future directions include a deeper exploration of the interplay between geometry and group mixing in dynamic or temporal settings, where connection patterns evolve over time.
Additionally, studying inference procedures for recovering RHBM parameters from real network data could unlock its potential for empirical applications, such as community-aware network embedding or generative modeling of attributed graphs.
We also plan to explore the use of the RHBM as a randomization tool for real networks, allowing one to preserve both latent geometry and meso-scale structure while perturbing finer topological details—thus enabling controlled null models for hypothesis testing.

While the model is referred to as \textit{hyperbolic} due to the isomorphism between the $\mathbb{S}^1$ and $\mathbb{H}^2$ models, the current formulation has been developed in Euclidean space for analytical convenience.
A natural next step is to explicitly formulate and analyze the RHBM directly in hyperbolic space, aligning its geometric underpinnings with the growing body of work in hyperbolic network science and potentially enhancing its descriptive power for hierarchical and tree-like structures.

\printbibliography

\end{document}